%% file: main.tex
\documentclass[runningheads]{llncs}
\usepackage[T1]{fontenc}
\usepackage{graphicx}
\usepackage{amsmath}
\usepackage{amsfonts}
\usepackage{amssymb}
\usepackage{booktabs}
\usepackage{calc}
\usepackage[sort]{cite}
\usepackage{hyperref}
\usepackage{listings}
\usepackage{mathtools}
\usepackage{stmaryrd}
\usepackage{subcaption}
\usepackage{tikz}
\usepackage{twoopt}
\usepackage{upgreek}
\usepackage{upquote}
\usepackage{xcolor}
\usepackage{xspace}

\input{commands}

\def\tr{} 
\newcommand{\iftr}[2]{\ifdefined\tr{#1}\else{#2}\fi}

\begin{document}
\title{Multiparty Session Typing, Embedded\iftr{\texorpdfstring{\\}{ }(Technical Report)}{}}
\author{Sung-Shik Jongmans\orcidID{0000-0002-4394-8745}}
\institute{University of Groningen, the Netherlands}
\maketitle
\begin{abstract}
Multiparty session typing (MPST) is a method to make concurrent programming
simpler. The idea is to use type checking to automatically detect safety and
liveness violations of implementations relative to specifications. In practice,
the premier approach to combine MPST with mainstream languages---in the absence
of native support---is based on \textit{external DSLs} and associated tooling.

In contrast, we study the question of how to support MPST by using
\textit{internal DSLs}. Answering this question positively, this paper presents
the \tool library: it leverages Scala's lightweight form of dependent typing,
called match types, to embed MPST directly into Scala. Our internal-DSL-based
approach avoids programming friction and leaky abstractions of the
external-DSL-based approach for MPST.

\end{abstract}


\section{Introduction}
\label{sect:intr}

\subsubsection*{Background}

With the advent of multicore processors, multithreaded program\-ming---a
notoriously error-prone en\-ter\-prise---has become increasingly important.

Because of this, mainstream languages have started to offer core support for
higher-level {\textit{communication} primitives} besides lower-level
{\textit{synchronisation} primitives} (e.g., Clojure, Go, Kotlin, Rust). The
idea has been to add \textit{message passing} as an abstraction for
\textit{shared memory}, as---supposedly---\textit{channels} are easier to use
than \textit{locks}. Yet, empirical research shows that ``message passing
does not necessarily make multithreaded programs less error-prone than shared
memory'' \cite{DBLP:conf/asplos/TuLSZ19}.

One of the core challenges is as follows: given a specification ${S}$ of the
\textit{communication protocols} that an implementation ${I}$ should fulfil, how
to prove that ${I}$ is \textit{safe} and \textit{live} relative to ${S}$? Safety
means ``bad'' communication actions never happen: \underline{if} a communication
action happens in ${I}$, \underline{then} it is allowed to happen by ${S}$.
Liveness means ``good'' communication actions eventually happen.

\subsubsection*{Multiparty session typing (MPST)}

MPST~\cite{DBLP:conf/popl/HondaYC08} is a method to automatically prove
safety and liveness of communication protocol implementations relative to
specifications. The idea is to write specifications as \emph{behavioural
types}~\cite{DBLP:journals/ftpl/AnconaBB0CDGGGH16,DBLP:journals/csur/HuttelLVCCDMPRT16} against which implementations are type-checked. Formally, the central theorem is that well-typed\-ness at com\-pile-time implies safety and liveness at run-time. Over the past 10--15 years, much progress has been made, including the development of many tools to combine MPST with mainstream languages (e.g., F\#~\cite{DBLP:conf/cc/NeykovaHYA18}, F$^\star$~\cite{DBLP:journals/pacmpl/00020HNY20}, Go~\cite{DBLP:journals/pacmpl/CastroHJNY19}, Java~\cite{DBLP:conf/fase/HuY16,DBLP:conf/fase/HuY17}, OCaml~\cite{DBLP:conf/ecoop/ImaiNYY19}, Rust~\cite{DBLP:conf/coordination/LagaillardieNY20,DBLP:conf/ecoop/LagaillardieNY22}, Scala~\cite{DBLP:conf/ecoop/ScalasDHY17,DBLP:conf/ecoop/CledouEJP22,DBLP:conf/issta/FerreiraJ23,DBLP:conf/ecoop/BarwellHY023}, and TypeScript~\cite{DBLP:conf/cc/Miu0Y021}).
\autoref{fig:mpst} visualises the idea behind MPST in more detail:

\begin{figure}[t]
	\begin{minipage}[b]{\linewidth*1/3}
		\begin{tikzpicture}[x=.5cm, y=-1.25cm, font=\footnotesize]
			\tikzstyle{object} = [inner sep=0mm, anchor=base, outer sep=1mm]
			\node [object] (G) at (0,0) {$G$};
			\node [object] (L1) at (-1.5,1) {$L_1$};
			\node [object] (L2) at (-.5,1) {$L_2$};
			\node [object] (Ldots) at (.5,1) {$\cdots$};
			\node [object] (Ln) at (1.5,1) {$L_n$};
			\node [object] (P1) at (-1.5,2) {$P_1$};
			\node [object] (P2) at (-.5,2) {$P_2$};
			\node [object] (Pdots) at (.5,2) {$\cdots$};
			\node [object] (Pn) at (1.5,2) {$P_n$};

			\tikzstyle{arrow} = [-stealth, rounded corners]
			\draw [arrow] (G.south) to (L1.north);
			\draw [arrow] (G.south) to (L2.north);
			\draw [arrow] (G.south) to (Ln.north);
			\draw [arrow] (L1.south) to (P1.north);
			\draw [arrow] (L2.south) to (P2.north);
			\draw [arrow] (Ln.south) to (Pn.north);

			\tikzstyle{label} = [inner sep=0mm, anchor=base east, minimum width=1.75cm]
			\node [label] at (-2.25,0) {global type};
			\node [label] at (-2.25,.5) {project};
			\node [label] at (-2.25,1) {local types};
			\node [label] at (-2.25,1.5) {type check};
			\node [label] at (-2.25,2) {processes};
		\end{tikzpicture}
		\caption{MPST}
		\label{fig:mpst}
	\end{minipage}%
	\begin{minipage}[b]{\linewidth*2/3}
		\tikzstyle{process} = [inner sep=.75mm, draw, anchor=south, rounded corners=.5mm, minimum height=3.75mm, minimum width=3.75mm, font=\scriptsize]
		\tikzstyle{lifeline} = [-|, dashed]
		\tikzstyle{com} = [ -latex]
		\tikzstyle{value} = [inner sep=0pt, anchor=base, yshift=.5mm, font=\scriptsize]
		\begin{minipage}{\linewidth*1/3}\centering
			\begin{tikzpicture}[x=2cm, y=-1cm*1/3, font=\footnotesize]
				\node [process] (a) at (0,-.75) {$\rolex[]{a}$};
				\node [process] (b) at (1,-.75) {$\rolex[]{b}$};
				\draw [lifeline, draw=none] (0,-.5) to (0,4.75);
				\draw [lifeline] (0,-.5) to (0,2.5);
				\draw [lifeline] (1,-.5) to (1,2.5);
				
				\draw [com] ([xshift=.5mm]0,0) to node [value] {$\datax[]{Propose(5)}$} ([xshift=-.5mm]1,0);
				\draw [com] ([xshift=-.5mm]1,1) to node [value] {$\datax[]{Accept}$} ([xshift=.5mm]0,1);
				\draw [com] ([xshift=.5mm]0,2) to node [value] {$\datax[]{Confirm}$} ([xshift=-.5mm]1,2);
			\end{tikzpicture}
			\subcaption{Run 1}
			\label{fig:abc-sd:a}
		\end{minipage}%
		\begin{minipage}{\linewidth*1/3}\centering
			\begin{tikzpicture}[x=2cm, y=-1cm*1/3, font=\footnotesize]
				\node [process] (a) at (0,-.75) {$\rolex[]{a}$};
				\node [process] (b) at (1,-.75) {$\rolex[]{b}$};
				\draw [lifeline, draw=none] (0,-.5) to (0,4.75);
				\draw [lifeline] (0,-.5) to (0,1.5);
				\draw [lifeline] (1,-.5) to (1,1.5);
				
				\draw [com] ([xshift=.5mm]0,0) to node [value] {$\datax[]{Propose(5)}$} ([xshift=-.5mm]1,0);
				\draw [com] ([xshift=-.5mm]1,1) to node [value] {$\datax[]{Reject}$} ([xshift=.5mm]0,1);
			\end{tikzpicture}
			\subcaption{Run 2}
			\label{fig:abc-sd:b}
		\end{minipage}%
		\begin{minipage}{\linewidth*1/3}\centering
			\begin{tikzpicture}[x=2cm, y=-1cm*1/3, font=\footnotesize]
				\node [process] (a) at (0,-.75) {$\rolex[]{a}$};
				\node [process] (b) at (1,-.75) {$\rolex[]{b}$};
				\draw [lifeline] (0,-.5) to (0,4.75);
				\draw [lifeline] (1,-.5) to (1,4.75);
				
				\draw [com] ([xshift=.5mm]0,0) to node [value] {$\datax[]{Propose(5)}$} ([xshift=-.5mm]1,0);
				\draw [com] ([xshift=-.5mm]1,1) to node [value] {$\datax[]{Propose(11)}$} ([xshift=.5mm]0,1);
				\draw [com] ([xshift=.5mm]0,2) to node [value] {$\datax[]{Propose(6)}$} ([xshift=-.5mm]1,2);
				\draw [com] ([xshift=-.5mm]1,3) to node [value] {$\datax[]{Propose(11)}$} ([xshift=.5mm]0,3);
				\draw [com] ([xshift=.5mm]0,4) to node [value] {$\datax[]{Reject}$} ([xshift=-.5mm]1,4);
			\end{tikzpicture}
			\subcaption{Run 3}
			\label{fig:abc-sd:c}
		\end{minipage}

		\caption{A few possible runs of the Negotation protocol}
		\label{fig:abc-sd}
	\end{minipage}
\end{figure}

\begin{enumerate}
	\item First, a protocol among roles $r_1, \ldots, r_n$ is implemented as a
	session of processes $P_1, \ldots, P_n$ (concrete), while it is specified as a
	\textit{global type} $G$ (abstract). The global type models the behaviour of
	all processes together.
	
	\item Next, $G$ is decomposed into local types $L_1, \ldots, L_n$ by
	\textit{projecting} $G$ onto each role. Each local type models the behaviour
	of one process alone.
	
	\item Last, safety and liveness are verified by
	\textit{type-checking} each $P_i$ against $L_i$.
\end{enumerate}

\begin{example}\label{exmp:nego}
	The \textit{Negotiation} protocol, originally defined in the MPST literature by
	Neykova et al. \cite{DBLP:conf/cc/NeykovaHYA18}, consists of roles
	\textit{Alice} and \textit{Bob}.
	\autoref{fig:abc-sd} shows three possible runs. First, a proposal is
	communicated from Alice to Bob. Next, its acceptance, rejection, or a
	counter-proposal is communicated from Bob to Alice. Next:
	
	\begin{itemize}
		\item In case of an acceptance, a confirmation is communicated from Alice to
		Bob.
		
		\item In case of a rejection, the protocol ends.
		
		\item In case of a counter-proposal, its acceptance, rejection, or another
		counter-proposal is communicated from Alice to Bob. And so on.
	\end{itemize}
	
	\noindent The following recursive \textbf{global type} specifies the protocol:
	\begin{center}$
		G = \begin{tree}
			\branchx{
				\rolex{a} \unbuf \rolex{b} \isa \datax{Propose(Int)} \pre \recx{X} \pre \rolex{b} \unbuf \rolex{a} \isa \holex{h1}
			}
			\branchxxx[h1]{
				\datax{Accept} \pre \rolex{a} \unbuf \rolex{b} \isa \datax{Confirm} \pre \one
			}{
				\datax{Reject} \pre \one
			}{
				\datax{Propose} \pre \rolex{a} \unbuf \rolex{b} \isa  \holex{h2}
			}
			\branchxxx[h2]{
				\datax{Accept} \pre \rolex{b} \unbuf \rolex{a} \isa \datax{Confirm} \pre \one
			}{
				\datax{Reject} \pre \one
			}{
				\datax{Propose(Int)} \pre X
			}
		\end{tree}
	$\end{center}
	Global type $p \unbuf q \isa \famxx{t_i \pre G_i}{1 \leq i \leq n}$ specifies
	the communication of a value of data type $t_i$ from role $p$ to role $q$,
	followed by $G_i$, for some
	${1}\thinspace{\leq}\thinspace{i}\thinspace{\leq}\thinspace{n}$; we omit braces
	when ${n}\thinspace{=}\thinspace{1}$. Global type $\one$ specifies termination.
	The following recursive \textbf{local type, projected from the global type,}
	specifies Bob (Alice is similar):
	\begin{center}$
		L_{\rolex[]{b}} = \begin{tree}
			\branchx{
				\rolex{a}\rolex{b} \recv \datax{Propose(Int)} \pre \recx{X} \pre \rolex{b}\rolex{a} \send \holex{h1}
			}
			\branchxxx[h1]{
				\datax{Accept} \pre \rolex{a}\rolex{b} \recv \datax{Confirm} \pre \one
			}{
				\datax{Reject} \pre \one
			}{
				\datax{Propose} \pre \rolex{a}\rolex{b} \recv  \holex{h2}
			}
			\branchxxx[h2]{
				\datax{Accept} \pre \rolex{b}\rolex{a} \send \datax{Confirm} \pre \one
			}{
				\datax{Reject} \pre \one
			}{
				\datax{Propose(Int)} \pre X
			}
		\end{tree}
	$\end{center}
	Local types $pq \send \famxx{t_i \pre L_i}{1 \leq i \leq n}$ and $pq \recv
	\famxx{t_i \pre L_i}{1 \leq i \leq n}$ specify the send and receive of a
	value of data type $t_i$ from role $p$ to role $q$, followed by $L_i$, for some
	${1}\thinspace{\leq}\thinspace{i}\thinspace{\leq}\thinspace{n}$; we omit braces
	when ${n}\thinspace{=}\thinspace{1}$. Local type $\one$ specifies termination.
	The following \textbf{process, well-typed by the local type,} implements a
	version of Bob:
	\begin{center}$\begin{aligned}
		P_{\rolex[]{b}} &= \datax{s} \atx{\rolex{a}\rolex{b}} \recv \datax{\_} \isa \datax{Propose(Int)} \prewide
	\\[-\baselineskip]
		 &\phantom{{}={}} \begin{tree}
			\branchx{
				\quad \loopx[\texttt{s}]{\datax{s} \atx{\rolex{a}\rolex{b}} \send \datax{propose(11)} \prewide \datax{s} \atx{\rolex{a}\rolex{b}} \recv \holex{h1}}
			}
			\branchxxx[h1]{
				\datax{\_} \isa \datax{Accept} \prewide \datax{s} \atx{\rolex{a}\rolex{b}} \send \datax{confirm} \prewide \nil
			}{
				\datax{\_} \isa \datax{Reject} \prewide \nil
			}{
				\datax{\_} \isa \datax{Propose(Int)} \prewide \recur[\texttt{s}]
			}
		\end{tree}
	\end{aligned}$\end{center}
	Process $x\atx{pq} \send e$ implements the send of the value of expression $e$
	from role $p$ to role $q$, followed by $P$, in session $x$. Process $x \atx{pq}
	\recv \famxx{x_i \isa t_i \pre P_i}{1 \leq i \leq n}$ implements the receive of
	a value of data type $t_i$ into variable $x_i$, followed by $P_i$, in session
	$x$, for some ${1}\thinspace{\leq}\thinspace{i}\thinspace{\leq}\thinspace{n}$;
	we omit braces when ${n}\thinspace{=}\thinspace{1}$. Process $\nil$ implements
	termination. Processes $\loopx[x]{P}$ and $\recur[x]$ implement iteration in
	session $x$. \qed
\end{example}

\begin{figure}[t]\centering
	\begin{tikzpicture}[x=2.5cm, font=\footnotesize]
		\tikzstyle{box} = [minimum width=1.5cm, minimum height=1.25cm*2/3, draw, align=center, rounded corners, outer sep=.5mm]
		\node [box] (Glob) at (0,0) {global\\[-.75mm]type};
		\node [box] (Locs) at (1,0) {local\\[-.75mm]types};
		\node [box] (DFAs) at (2,0) {FSMs};
		\node [box] (APIs) at (3,0) {APIs};
		\node [box] (Procs) at (4,0) {\smash{p}rocesses};
		
		\draw [-latex] (Glob) to node [label, above=1mm, align=center] {project\\(auto)} (Locs);
		\draw [-latex] (Locs) to node [label, above=1mm, align=center] {interpret\\(auto)} (DFAs);
		\draw [-latex] (DFAs) to node [label, above=1mm, align=center] {encode\\(auto)} (APIs);
		\draw [-latex] (APIs) to node [label, above=1mm, align=center] {use\\(man)} (Procs);
		
		\draw [decorate, decoration={brace, mirror, amplitude=2mm}] ([yshift=-1.25mm]Glob.south west) to node [anchor=north, yshift=-2mm] {\clap{\textit{external DSL}}} ([yshift=-1.25mm]Glob.south east);
		\draw [decorate, decoration={brace, mirror, amplitude=2mm}] ([yshift=-1.25mm]APIs.south west) to node [anchor=north, yshift=-2mm, align=center] {\textit{mainstream} \textit{language}} ([yshift=-1.25mm]Procs.south east);
	\end{tikzpicture}
	
	\caption{Workflow of external-DSL-based MPST tools}
	\label{fig:apigen}
\end{figure}
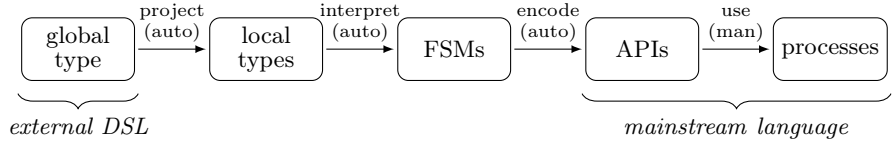

In practice \cite{DBLP:books/sp/24/Yoshida24}, the premier approach to combine
MPST with mainstream languages\allowbreak---in the absence of native
support---is based on: (1) external DSLs\footnote{A \textit{domain-specific
language} (DSL) is either \textit{external} or \textit{internal}. External DSLs
are stand-alone languages with their own dedicated syntax, while internal DSLs
are embedded languages into a \textit{general-purpose language} (GPL) with
syntax inherited from that GPL. Both approaches have advantages and
disadvantages \cite{DBLP:books/daglib/0034522}.} to write global types; (2)
associated tooling to generate corresponding code in mainstream languages,
including Scribble~\cite{DBLP:conf/fase/HuY16,DBLP:conf/fase/HuY17}, its
extensions~\cite{DBLP:conf/ecoop/ScalasDHY17,DBLP:journals/pacmpl/CastroHJNY19,DBLP:conf/cc/Miu0Y021,DBLP:conf/coordination/LagaillardieNY20,DBLP:conf/coordination/CutnerY21,DBLP:conf/ecoop/LagaillardieNY22,DBLP:conf/ppopp/CutnerYV22,DBLP:conf/cc/NeykovaHYA18,DBLP:journals/pacmpl/00020HNY20}, StMungo~\cite{DBLP:journals/scp/KouzapasDPG18}, {\small\texttt{mpstpp}}~\cite{DBLP:conf/esop/JongmansY20}, $\upnu$Scr~\cite{DBLP:conf/fct/YoshidaZF21}, Pompset~\cite{DBLP:conf/ecoop/CledouEJP22}, Teatrino~\cite{DBLP:conf/ecoop/BarwellHY023}, and Oven~\cite{DBLP:conf/issta/FerreiraJ23}.

The key ideas of the external-DSL-based approach were originally conceived
by Deni{\'{e}}lou, Hu, and Yoshida. It is based on two insights: local types can
be interpreted as \textit{finite-state machines}
(FSM)~\cite{DBLP:conf/esop/DenielouY12,DBLP:conf/icalp/DenielouY13}, where
states and transitions model sends and receives; FSMs can be encoded as
object-oriented \emph{application programming interfaces}
(API)~\cite{DBLP:conf/fase/HuY16,DBLP:conf/fase/HuY17}, where classes and
methods model states and transitions. \autoref{fig:apigen} visualises the
workflow. First, the programmer writes a global type in a DSL; this is the input
of the MPST tool. Next, the MPST tool projects the global type to local types,
interprets the local types as FSMs, and encodes the FSMs as APIs in the
mainstream language; this is the output of the MPST tool. Last, the programmer
uses the APIs to write processes.

\begin{figure}[t]
	\begin{minipage}[b]{\linewidth*4/7-2mm}
		\begin{lstlisting}[language=scrib]
Propose from A to B; rec X {
  choice at B
      { Accept  from B to A;
        Confirm from A to B; }
   or { Reject  from B to A; }
   or { Propose from B to A;
        choice at A
            { Accept  from A to B;
              Confirm from B to A; }
         or { Reject  from A to B; }
         or { Propose from A to B;
              continue X; } } } $\tikz[overlay, remember picture]{\node [inner sep=0pt, circle, anchor=base] (x) {\phantom{\{}};}$
		\end{lstlisting}%
		\caption{Global type for Negotation (Scribble)}
		\label{fig:nego-scrib}
	\end{minipage}%
	\hfill%
	\begin{minipage}[b]{\linewidth*3/7-2mm}\centering
		\begin{tikzpicture}[x=1.125cm, y=1.125cm, inner sep=0pt, font=\scriptsize]
			\tikzstyle{state} = [minimum width=4mm, draw, circle];
			\tikzstyle{final} = [minimum width=3mm, draw, circle];
			\tikzstyle{trans} = [->];
			\node [state] (s1) at ($(0,0)+(-150:1)+(90:1)$) {1};
			\node [state] (s2) at ($(0,0)+(-150:1)$) {2};
			\node [state] (s3) at ($(0,0)+(-30:1)$) {3};
			\node [state] (s4) at ($(s2)+(-90:1)+(-150:1)$) {4};
			\node [state] (s5) at ($(s2)+(-90:1)+(-30:1)$) {5};
			\node [state] (s6) at ($(s3)+(-90:1)+(-30:1)$) {6};
			\node [final] at (s5) {};
			\draw [trans] ([xshift=-1cm*1/3]s1.west) to (s1);
			\draw [trans] (s1) to node [left=.5mm] {$\datax[]{AB} \recv \datax[]{Propose}$} (s2);
			\draw [trans] (s2) to node [sloped, below=.5mm] {$\datax[]{BA} \send \datax[]{Propose}$} (s3);
			\draw [trans] (s2) to node [sloped, above=.5mm] {$\datax[]{BA} \send \datax[]{Accept}$} (s4);
			\draw [trans] (s2) to node [sloped, below=.5mm] {$\datax[]{BA} \send \datax[]{Reject}$} (s5);
			\draw [trans, out=135, in=45] (s3) to node [pos=1/2, sloped, above=.5mm] {$\datax[]{AB} \recv \datax[]{Propose}$} (s2);
			\draw [trans] (s3) to node [sloped, above=.5mm] {$\datax[]{AB} \recv \datax[]{Reject}$} (s5);
			\draw [trans] (s3) to node [sloped, below=.5mm] {$\datax[]{AB} \recv \datax[]{Accept}$} (s6);
			\draw [trans] (s4) to node [sloped, below=.5mm] {$\datax[]{AB} \recv \datax[]{Confirm}$} (s5);
			\draw [trans] (s6) to node [sloped, above=.5mm] {$\datax[]{BA} \send \datax[]{Confirm}$} (s5);
		\end{tikzpicture}
		\caption{FSM for Bob}
		\label{fig:nego-fsm}
	\end{minipage}

	\bigbreak

	\begin{minipage}[b]{\linewidth*3/5-2mm}\centering%
		\begin{lstlisting}[language=scala, numbers=none]
trait Loop[S]:
  def loop(f: ((S => S5, S) => S5)): S5

trait S1:
  def recvFromA(f: (Propose, S2) => S5): S5

trait S2 extends Loop[S2]:
  def sendToA(v: Accept,  f: S4 => S5): S5
  def sendToA(v: Reject,  f: S5 => S5): S5
  def sendToA(v: Propose, f: S3 => S5): S5

trait S3 extends Loop[S3]:
  def recvFromA(f1: (Accept,  S6) => S5,
                f2: (Reject,  S5) => S5,
                f3: (Propose, S2) => S5): S5

... // traits S4, S5, and S6
		\end{lstlisting}
		\caption{Callback-based API for Bob (Scala)}
		\label{fig:nego-api}
	\end{minipage}%
	\hfill
	\begin{minipage}[b]{\linewidth*2/5-2mm}\centering%
		\begin{lstlisting}[language=scala]
class Propose(val x: Int)
class Accept
class Reject
class Confirm

val v = new Propose(11)

def bob(s1: S1): S5 =
  s1.recvFromA((x, s2) =>
    s2.loop((recur, s2) =>
      s2.sendToA(v, s3 =>
        s3.recvFromA(
          (_, s6) =>
            s6.sendToA(...),
          (_, s5) => s5,
          (_, s2) =>
            recur(s2)))))
		\end{lstlisting}
		\caption{Process for Bob}
		\label{fig:nego-proc}
	\end{minipage}%
\end{figure}

\begin{example}\label{exmp:nego2}
	\autoref{fig:nego-scrib} shows a global type for Negotiation (cf. $G$ in
	\autoref{exmp:nego}), written in the external DSL of Scribble. %
	Statement \scribm!$t$ from $p$ to $q$! specifies the communication of a value
	of data type $t$ from $p$ to $q$. Statement %
	\scribm!choice at $r$ { $G_1$!\allowbreak\scribm!} or $\cdots$ or { $G_k$ }!
	specifies a choice among $G_1, \ldots, G_k$ made by $r$.
	
	\autoref{fig:nego-fsm} shows the FSM for Bob, derivable from
	\autoref{fig:nego-scrib}. Transition labels $pq \send t$ and $pq \recv t$
	specify the send and receive of a value of data type $t$ from $p$ to $q$.
	
	\autoref{fig:nego-api} shows a \textit{callback-based API} for Bob in Scala,
	derivable from \autoref{fig:nego-fsm}. Trait \scalam!S$i$! in the API
	corresponds with state $i$ of the FSM; methods of trait \scalam!S$i$!
	correspond with transitions of state $i$. Traits \scalam!S2! and \scala!S3!
	also extend trait \scala!Loop! to be able to start callback-based iteration in
	states $2$ and $3$ (i.e., these are the only states on a cycle in the FSM) in a
	type-sound manner. We note that each method and each callback returns a value
	of type \scala!S5! to ensure that the program can terminate only when the final
	state has been reached.
	
	To demonstrate the usage of the API, \autoref{fig:nego-proc} shows a process
	for Bob (cf. $P$ in \autoref{exmp:nego}). The idea is to write a function,
	\scala!bob!, that consumes an ``initial state object'' \scala!s1! as input and
	produces a ``final state object'' \scala!s5! as output. First, the only
	communication action that can be performed, is the one for which \scala!s1! has
	a method (receiving). When that method is called, the actual receive is
	performed, and the callback is called with the received value \scala!x! and a
	fresh ``successor state object'' \scala!s2!. Next, the only communication
	actions that can be performed, are the ones for which \scala!s2! has a method
	(sending). And so on. \qed
\end{example}

\subsubsection*{This work}

The external-DSL-based approach is well-established in the MPST literature: it
is used in all MPST tools
\cite{DBLP:conf/fase/HuY16,DBLP:conf/fase/HuY17,DBLP:conf/ecoop/ScalasDHY17,DBLP:journals/pacmpl/CastroHJNY19,DBLP:conf/cc/Miu0Y021,DBLP:conf/coordination/LagaillardieNY20,DBLP:conf/coordination/CutnerY21,DBLP:conf/ecoop/LagaillardieNY22,DBLP:conf/ppopp/CutnerYV22,DBLP:conf/cc/NeykovaHYA18,DBLP:journals/pacmpl/00020HNY20,DBLP:journals/scp/KouzapasDPG18,DBLP:conf/esop/JongmansY20,DBLP:conf/fct/YoshidaZF21,DBLP:conf/ecoop/CledouEJP22,DBLP:conf/issta/FerreiraJ23,DBLP:conf/ecoop/ImaiNYY19} that support \textit{classical} MPST as in \autoref{fig:mpst} (global types and projection; fully automatic; static up-to linearity). However, despite the major impact, it has two weaknesses:
\begin{description}
	\item[Programming friction:] The usage of an external DSL to specify protocols
	as global types causes programming friction. In general, this is a
	well-document\-ed issue with external DSLs (e.g.,
	\cite{DBLP:books/daglib/0034522}): new syntax needs to be learned; new tools to
	edit DSL code need to be adopted; extra effort is needed to intermix DSL code
	with the mainstream language.
	
	\item[Leaky abstractions:] As demonstrated in \autoref{exmp:nego2}, APIs
	generated by MPST tools leak internal details: global types are essentially
	declarative, whereas the FSMs that seep through the APIs are essentially
	imperative. This representational gap causes dissonance between the level of
	abstraction at which global types are produced by the programmer (before API
	generation), and the level of abstraction at which local types are consumed by
	that same programmer \textit{in terms of FSMs} (after API generation).
\end{description}

To avoid these weaknesses, we explore a different approach and study the
question of how to support classical MPST by using \textit{internal DSLs}.
Answering this question positively, we present the \tool library: it leverages
Scala's ``lightweight form of dependent typing''
\cite{DBLP:journals/pacmpl/BlanvillainBKO22}, called \textit{match types}, to
embed global\slash local types directly into Scala. As a result, \tool offers a
frictionless interface between global\slash local types and processes (i.e., no
new syntax, editors, or other tools need to be adopted). Moreover, \tool avoids
leaky abstractions by not relying on FSMs; global\slash local types are
first-class citizens.

In this way, \tool is the first internal-DSL-based MPST tool that supports all
key aspects of classical MPST as in \autoref{fig:mpst} (unlike Imai et al.
\cite{DBLP:conf/ecoop/ImaiNYY19}, who do not support $n$-ary choice and require
extra manual work to guide projection). This is a significant contribution,
because: (a) internal DSLs have advantages over external DSLs, but (b) it is far
from obvious how to build an internal DSL for MPST in a mainstream language
without native support for session types.

Technically, to apply classical MPST and offer static guarantees, some form of
compile-time computation is needed. This is the role of match types. They are
essentially match expressions at the type level, which are evaluated by the
Scala compiler as part of its static analysis, and which we use in this work to
embed MPST theory. That is, the Scala compiler can check the typing rules of
MPST theory by evaluating carefully crafted match types.

First, through extensive examples, we give an overview of the capabilities of
\tool (\autoref{sect:overv:basics} and \autoref{sect:overv:merge}). Next, we
present technical details (\autoref{sect:detail}). Last, we conclude this paper
with related work and future work (\autoref{sect:concl}).


\section{A Tour of \texttt{mpst.embedded}: Basic Features}
\label{sect:overv:basics}

\subsubsection*{Global types}

\begin{figure}[t]\footnotesize\centering
	\begin{tabular}{@{} l @{\quad} l @{}}
		\toprule
		\textbf{Global types:}
	\\	\ \scalam!Com[$p$, $q$, (($t_1$, $G_1$), $\ldots$, ($t_n$, $G_n$))]!, & $p \unbuf q \isa \famxx{t_i \pre G_i}{1 \leq i \leq n}$,
	\\	\ \scala!End!, \scalam!Loop[$X$, $G$]!, \scalam!Recur[$X$]! & $\one$, $\recx{X} \pre G$, $X$
	\\	
		\textbf{Local types:}
	\\	\ \scalam!Send[$p$, $q$, (($t_1$, $L_1$), $\ldots$, ($t_n$, $L_n$))]!, & $pq \send \famxx{t_i \pre L_i}{1 \leq i \leq n}$,
	\\	\ \scalam!Recv[$p$, $q$, (($t_1$, $L_1$), $\ldots$, ($t_n$, $L_n$))]!, & $pq \recv \famxx{t_i \pre L_i}{1 \leq i \leq n}$,
	\\	\ \scala!End!, \scalam!Loop[$X$, $L$]!, \scalam!Recur[$X$]! & $\one$, $\recx{X} \pre L$, $X$
	\\	
		\textbf{Processes}
	\\	\ \scalam!$x$.send($q$, $e$, (_) => $P$)!, & $x\atx{pq} \send e$,
	\\	\ \scalam!$x$.recv($p$, (($x_1$: $t_1$, _) => $P_1$, $\ldots$, ($x_n$: $t_n$, _) => $P_n$))!, & $x \atx{pq} \recv \famxx{x_i \isa t_i \pre P_i}{1 \leq i \leq n}$,
	\\	\ \scalam!$x$.loop((recur, _) => $P$)!, \scalam!recur($x$)! & $\loopx[x]{P}$, $\recur[x]$
	\\	\bottomrule
	\end{tabular}
	
	\caption{Correspondence between \tool (left) and MPST theory (right)}
	\label{fig:relation}
\end{figure}

\autoref{fig:relation} (top rows) shows the correspondence between global types
in \tool and in MPST theory. In \tool, each global type $G$ is built from
classes \scala!Com!, \scala!End!, \scala!Loop!, and \scala!Recur!. The third
type parameter of \scala!Com! is an $n$-ary product type, called ``the
branches''. Type parameter $X$ of \scala!Loop! is bound in type parameter $G$ to
the whole \scalam!Loop[$X$, $G$]!, to embed a recursive type. Each role $p$ or
$q$, and each recursion variable $X$, is a Scala \textit{string literal type}
(e.g., \scalatype!"foo"! is a type with one inhabitant, \scala!"foo"!). Each
data type $t$ is a Scala type.

\begin{figure}[p]
	\begin{minipage}{\linewidth}
		\begin{lstlisting}[language=scala, numbers=left]
type S =
  Com["A", "B", ((Propose,
    Loop["X",
      Com["B", "A", (
        (Accept,  Com["A", "B", ((Confirm, End))]),
        (Reject,  End),
        (Propose, Com["A", "B", (
          (Accept,  Com["B", "A", ((Confirm, End))]),
          (Reject,  End),
          (Propose, Recur["X"]))]))]]))]
		\end{lstlisting}
		\caption{Global type for Negotiation}
		\label{fig:nego-glob-mpst4}
	\end{minipage}
	
	\medbreak
	
	\begin{minipage}{\linewidth}
		\begin{lstlisting}[language=scala, numbers=left]
type `S@B` = // equivalent to Proj[S, "B"] -- S is defined in Fig. $\tt\color{gray}\ref{fig:nego-glob-mpst4}$
  Recv["A", "B", ((Propose,
    Loop["X",
      Send["B", "A", (
        (Accept,  Recv["A", "B", ((Confirm, End))]),
        (Reject,  End),
        (Propose, Recv["A", "B", (
          (Accept,  Send["B", "A", ((Confirm, End))]),
          (Reject,  End),
          (Propose, Recur["X"]))]))]]))]
		\end{lstlisting}
		\caption{Local type for Bob}
		\label{fig:nego-loc-bob-mpst4}
	\end{minipage}
	
	\medbreak
	
	\begin{minipage}{\linewidth}
		\begin{lstlisting}[language=scala, numbers=left]
def alice(
    s: Local["A", Proj[S, "A"]] // S is defined in Fig. $\tt\color{gray}\ref{fig:nego-glob-mpst4}$
    ): Local["A", End] =
  s.send("B", new Propose(5), s => 
    s.recv("B", (
      (_, s) => s.send("B", new Confirm, s => s),
      (_, s) => s,
      (v, s) =>
        if 
          v.x < 11
        then 
          s.send("B", new Accept, s =>
            s.recv("B", (_, s) => s))
        else 
          s.send("B", new Propose(6), s =>
            s.recv("B", (
              (_, s) => s.send("B", new Confirm, s => s),
              (_, s) => s,
              (_, s) => s.send("B", new Reject, s => s)))))))

def bob(
    s: Local["B", `S@B`] // `S@B` is defined in Fig. $\tt\color{gray}\ref{fig:nego-loc-bob-mpst4}$
    ): Local["B", End] =
  s.recv("A", (_, s) =>
    s.loop((recur, s) =>
      // val error = s // redundant line -- only used in Exmp. $\tt\color{gray}\ref{exmp:errors}$
      s.send("A", new Propose(11), s =>
        s.recv("A", (
          (_, s) => s.send("A", new Confirm, s => s),
          (_, s) => s,
          (_, s) => recur(s))))))

val s = new Global[S]
val _ = new Thread(() => { alice(s.init["A"]); () }).start
val _ = new Thread(() => { bob  (s.init["B"]); () }).start
		\end{lstlisting}
		\caption{Processes for Alice and Bob}
		\label{fig:nego-proc-bob-mpst4}
	\end{minipage}
\end{figure}

\begin{example}
	\autoref{fig:nego-glob-mpst4} shows a global type for Negotiation (cf. $G$
	in \autoref{exmp:nego}). \qed
\end{example}

\subsubsection*{Local types and projection}

\autoref{fig:relation} (middle rows) shows the correspondence between local
types in \tool and in MPST theory. We add that local types can be computed from
global types fully automatically and statically via type \scala!Proj!: the Scala
compiler reduces \scalam!Proj[$G$, $r$]! to the projection of $G$ onto $r$.

\begin{example}
	\autoref{fig:nego-loc-bob-mpst4} shows a local type for Bob (cf.
	$L_{\rolex[]{b}}$ in \autoref{exmp:nego}). Alternatively, it can be computed by
	having the Scala compiler reduce \scala!Proj[S, "B"]!. \qed
\end{example}

\subsubsection*{Processes and type checking}

\autoref{fig:relation} (bottom rows) shows the correspondence between processes
in \tool and in MPST theory. In \tool, each process is a sequence of calls to
methods \scala!send!, \scala!recv!, \scala!loop!, and \scala!recur! of class
\scala!Local!. This generic class has two type parameters: one to represent a
role (enacted by the process), and another one to represent a local type (with
which the process must comply). In turn, instances of \scala!Local! are obtained
through calls to method \scala!init! of class \scala!Global!. This generic class
has one type parameter to represent a global type (with which all processes must
comply). Method \scala!init! consumes a role, initialises the session for it,
and produces a \scala!Local! object for it. Calls to \scala!init! are
\textit{blocking}: they return only when \textit{all} processes have called
\scala!init!.

Intuitively, \scala!Global! and \scala!Local! objects represent executable
sessions from the global and local perspective, leveraging the same abstractions
as the global and local types by which they are parametrised (no leaky
abstractions).

\begin{example}\label{exmp:nego3}
	\autoref{fig:nego-proc-bob-mpst4} shows processes for Alice and Bob on lines
	1--19 and 21--31 (cf. $P$ in \autoref{exmp:nego}), plus session initiation on
	lines 33--35. We make three remarks:
	\begin{itemize}
		\item The process for Bob looks similar to \autoref{fig:nego-proc}. However,
		\autoref{fig:nego-proc-bob-mpst4} is defined in terms of communication actions
		in a session (\scala!Local! objects), whereas \autoref{fig:nego-proc} is
		defined in terms of transitions of an FSM (\scalam!S$i$! objects).
		
		\item The process for Bob exactly mimics the recursive structure of local type
		\scala!`S@B`!. Such mimicry is not a general requirement for well-typed
		processes, as demonstrated by the process for Alice: instead of exactly
		mimicking the recursive structure of \scala!Proj[S, "A"]! (which has a similar
		recursive structure as \scala!S! in \autoref{fig:nego-glob-mpst4}), it mimics
		two \textit{unfoldings} of \scala!Proj[S, "A"]!, followed by termination. That
		is, lines 5--15 in \autoref{fig:nego-proc-bob-mpst4} comply with the first
		unfolding, while lines 16--19 comply with the second unfolding, without
		entering a loop. \qed
	\end{itemize}
\end{example}

Using \tool, the Scala compiler statically checks for each call $\alpha$ on a
\scala!Local! object, parametrised by local type $L$, whether or not $\alpha$
complies with $L$. If not, the Scala compiler reports an error. In this way,
\tool assures that well-typedness at compile-time implies safety and liveness at
run-time, modulo linear usage of \scala!Local! objects (checked dynamically),
and modulo non-terminating\slash ex\-cep\-tion\-al behaviour (unchecked). These
\textbf{two provisos}\label{provisos} are standard for MPST tools. As the type
parameters of \scala!Local! objects are erased at compile-time, only generic
\scala!Local! objects exist at run-time.

\begin{example}\label{exmp:errors}
	The following protocol violations are reported at compile-time:
	\begin{itemize}
		\item In \autoref{fig:nego-proc-bob-mpst4}, replace line 29  with one of the following:
		
		\smallbreak
			\begin{lstlisting}[language=scala]
(_, s) => s.send("A", new Reject, s => s),  // wrong data type
			\end{lstlisting}
		\smallbreak
			\begin{lstlisting}[language=scala]
(_, s) => s.send("C", new Confirm, s => s), // wrong receiver
			\end{lstlisting}
		\smallbreak
			\begin{lstlisting}[language=scala]
(_, s) => s.recv("A", (_, s) => s),         // wrong communication action
			\end{lstlisting}
		\smallbreak
		
		\item In \autoref{fig:nego-proc-bob-mpst4}, uncomment line 26 and replace line
		31 with:
		
		\smallbreak
			\begin{lstlisting}[language=scala]
(_, s) => recur(error))))))                 // wrong recursive type
			\end{lstlisting}
	\end{itemize}
	The following protocol violation is reported as an error at run-time:
	\begin{itemize}
		\item In \autoref{fig:nego-proc-bob-mpst4}, replace line 31 with:
		
		\smallbreak
			\begin{lstlisting}[language=scala]
(_, s) => { recur(s); recur(s) })))))       // linearity violation
			\end{lstlisting}
	\end{itemize}
	The \iftr{appendix (\autoref{sect:screen})}{technical report \cite{techreport}}
	contains a screenshot of error reporting. \qed
\end{example}

Besides protocol violations, additionally, basic well-formedness violations of
global types are reported as errors at compile-time; they are checked as part of
the instantiation of generic class \scala!Global! (e.g.,
\autoref{fig:nego-proc-bob-mpst4}, line 33). For instance, for %
\scalam!Com[$p$, $q$, (($t_1$, $G_1$), $\ldots$, ($t_n$, $G_n$))]!, we always
require $p \neq q$ and $n \geq 1$.

\section{The Tour, Continued: Advanced Features}
\label{sect:overv:merge}

\subsubsection*{Full merging}

To project global types, an auxiliary partial operator to \textit{merge} local
types---the projections---is needed. There are two variants
\cite{DBLP:journals/pacmpl/ScalasY19}: ``plain'' (basic) and ``full''
(advanced). Plain merge is \textit{relatively easy} to support, but it works for
\textit{few} local types, so many global types cannot be projected. Conversely,
full merge works for \textit{many} local types, but it is \textit{relatively
hard} to support. For instance, Imai et al. \cite{DBLP:conf/ecoop/ImaiNYY19}
support only manual full merge (i.e., the programmer must write extra
protocol-specific code to guide the computation of projections). In contrast,
\tool supports automatic full merge via type \scala!Merg!: the Scala compiler
reduces %
\scalam!Merg[$L_1$, $L_2$]! to the full merge of $L_1$ and $L_2$.

\begin{figure}[t]
	\begin{minipage}{\linewidth*1/2-2mm}
		\begin{lstlisting}[language=scala, numbers=left]
type S =
  Com["B1", "S", ((String,
    Com["S", "B1", ((Int,
      Com["S", "B2", ((Int,
        Com["B1", "B2", ((Int,
          T))]))]))]))]

type T =
  Com["B2", "B1", (
    (Ok,
      Com["B2", "S", ((Ok,
        Com["B2", "S", ((String,
          Com["S", "B2", ((Date,
            End))]))]))]),
    (Quit,
      Com["B2", "S", ((Quit,
        End))]))]
		\end{lstlisting}
		\caption{Global type for Two-Buyer}
		\label{fig:two-buyer-glob-mpst4}
	\end{minipage}%
	\hfill%
	\begin{minipage}{\linewidth*1/2-2mm}
		\begin{lstlisting}[language=scala, numbers=left]
type `S@S` = // equiv. Proj[S, "S"]
  Recv["B1", "S", ((String,
    Send["S", "B1", ((Int,
      Send["S", "B2", ((Int,
        // ignore Int from B1 to B2
          `T@S`))]))]))]))]

type `T@S` = // equiv. Proj[T, "S"]
  Merg[(
    // ignore Ok from B2 to B1
      Recv["B2", "S", ((Ok,
        Recv["B2", "S", ((String,
          Send["S", "B2", ((Date,
            End))]))]))],
    // ignore Quit from B2 to B1
      Recv["B2", "S", ((Quit,
        End))])]
		\end{lstlisting}
		\caption{Local type for Seller}
		\label{fig:two-buyer-loc-seller-mpst4}
	\end{minipage}
	
	\medbreak
	
	\begin{minipage}{\linewidth*1/2-2mm}
		\begin{lstlisting}[language=scala, numbers=left]
def seller(
    s: Local["S", Proj[S, "S"]]
    ): Local["S", End] =
  s.recv("B1", (_, s) =>
    val v = 11
    s.send("B1", v, s =>
      s.send("B2", v, s =>
        s.recv("B2", (
          (_: Ok, s) => 
            s.recv("B2", (_, s) =>
              val q: "B2" = "B2"
              val v = new Date
              s.send(q, v, s => s)),
          (_: Quit, s) => s)))))
		\end{lstlisting}
		\caption{Process for Seller}
		\label{fig:two-buyer-proc-seller-mpst4}
	\end{minipage}%
	\hfill
	\begin{minipage}{\linewidth*1/2-2mm}
		\begin{lstlisting}[language=scala, numbers=left]
// one session between B1, B2, and S
type S = ... // Fig. $\tt\color{gray}\ref{fig:two-buyer-glob-mpst4}$
type T = ... // Fig. $\tt\color{gray}\ref{fig:two-buyer-glob-mpst4}$

// another session between B2 and B3
type U =
  Com["B2", "B3", ((Int,
    Com["B2", "B3", ((Delegatee,
      Com["B3", "B2", (
        (Ok, End),
        (Quit, End))]))]))]

type Delegatee =
  Local["B2", Proj[T, "B2"]]
		\end{lstlisting}
		\caption{Global types for Three-Buyer}
		\label{fig:three-buyer-glob-mpst4}
	\end{minipage}
	
	\medbreak
	
	\begin{minipage}[t]{\linewidth*1/2-2mm}
		\begin{lstlisting}[language=scala, numbers=left]
def buyer2( // Three-Buyer version
    s: Local["B2", Proj[S, "B2"]],
    u: Local["B2", Proj[U, "B2"]]
    ): Local["B2", End] =
  s.recv("S", (x, s) =>
    s.recv("B1", (y, s) =>
      u.send("B3", x - y, u =>
        u.send("B3", s, u =>
          u.recv("B3", (
            (_, u) => u, ...))))))
		\end{lstlisting}
		\caption{Process for Buyer2}
		\label{fig:three-buyer-proc-buyer2-mpst4}
	\end{minipage}%
	\hfill%
	\begin{minipage}[t]{\linewidth*1/2-2mm}
		\begin{lstlisting}[language=scala, numbers=left]
def buyer3(
    u: Local["B3", Proj[U, "B3"]]
    ): Local["B3", End] =
  u.recv("B2", (_, u) =>
    u.recv("B2", (s, u) =>
      val v = new Quit
      u.send("B2", v, u =>
        s.send("B1", v, s =>
          s.send("S", v, s => s))
        u)))
		\end{lstlisting}
		\caption{Process for Buyer3}
		\label{fig:three-buyer-proc-buyer3-mpst4}
	\end{minipage}
\end{figure}

\begin{example}\label{exmp:two-buyer}
	The \textit{Two-Buyer} protocol, originally defined in the MPST literature by
	Honda et al. \cite{DBLP:conf/popl/HondaYC08}, consists of roles
	\textit{Buyer1}, \textit{Buyer2}, and \textit{Seller}: ``[Buyer1 and Buyer2]
	wish to buy an expensive book from Seller by combining their money. Buyer1
	sends the title of the book to Seller, Seller sends to both Buyer1 and Buyer2
	its quote, Buyer1 tells Buyer2 how much she can pay, and Buyer2 either accepts
	the quote or rejects the quote by notifying Seller.'' We use an extended
	version defined by Coppo et al. \cite{DBLP:journals/mscs/CoppoDYP16}, in which
	Buyer2 notifies not only Seller about acceptance\slash rejection, but also
	Buyer1. In the case of acceptance, Buyer2 sends his address to Seller, and
	Seller sends back the delivery date to Buyer2.
	
	\autoref{fig:two-buyer-glob-mpst4} and \autoref{fig:two-buyer-loc-seller-mpst4}
	show a global type for Two-Buyer and a local type for Seller (split into
	\scala!S!/\scala!S@S! and \scala!T!/\scala!T@S! for presentational reasons.)
	First, we note that the communication from Buyer1 to Buyer2 on line 5 in the
	global type has no counterpart on line 5 in the local type; as Seller does not
	participate in the communication, it is simply skipped in the projection.
	Second, we note that the communication from Buyer2 to Buyer1 on line 9 in the
	global type is ignored, too, but the projections of the two branches do need to
	be combined into one. This is achieved by having the Scala compiler reduce
	\begin{center}
	\scala!Merg[Recv["B2", "S", ((Ok, ...))], Recv["B2", "S", ((Quit, ...))]]!
	\end{center}
	to \scala!Recv["B2", "S", ((Ok, ...), (Quit, ...))]!.
	
	\autoref{fig:two-buyer-proc-seller-mpst4} shows a process for Seller. It
	demonstrates that merging is a type-level concept, hidden from the programmer:
	the Scala compiler reduces \scala!Merg[...]! and type-checks the code against
	the result transparently. \qed
\end{example}

\subsubsection*{Delegation}

Sessions are \textit{higher-order}: \scala!Local! object \scala!s! for a first
session can be \textit{delegated} between processes via \scala!Local! object
\scala!u! for a second session, by sending \scala!s! via \scala!u!. In the
presence of delegation, within each session, well-typedness at compile-time
continues to imply safety and liveness at run-time (modulo the ``two provisos'';
\autopageref{provisos}). However, between sessions, liveness is not assured;
supporting this would require substantial extra technical machinery
\cite{DBLP:journals/mscs/CoppoDYP16}, so none of the existing MPST tools support
it.

\begin{example}\label{exmp:three-buyer}
	The \textit{Three-Buyer} protocol, originally defined in the MPST literature by
	Coppo et al. \cite{DBLP:journals/mscs/CoppoDYP16}, consists of roles
	\textit{Buyer1}, \textit{Buyer2}, \textit{Buyer3}, and \textit{Seller}. It
	resembles the Two-Buyer protocol, except that Buyer2 can ask Buyer3 to enact
	his role on his behalf---unbeknownst to Buyer1 and Seller---through delegation.
	
	\autoref{fig:three-buyer-glob-mpst4} shows global types for Three-Buyer. Global
	type \scala!S! specifies the first sub-protocol among Buyer1, Buyer2, and
	Seller; it is identical to the global type for Two-Buyer. Global type \scala!U!
	specifies the second sub-protocol between Buyer2 and Buyer3. Notably, line 8
	specifies the delegation from Buyer2 to Buyer3.
	
	\autoref{fig:three-buyer-proc-buyer2-mpst4} shows a process for Buyer2: on
	lines 1--4, \scala!Local! objects for two sessions are consumed as inputs (to
	engage in two sub-protocols); on lines 5--6, the first session is used; on
	lines 7--10, the second session is used; on line 8, the remainder of the first
	session is delegated via the second session. Similarly,
	\autoref{fig:three-buyer-proc-buyer3-mpst4} shows a process for Buyer3: on
	lines 1--3, a \scala!Local! object for the first session is consumed as input;
	on line 5, a \scala!Local! object for the second session is received.
	
	The process for Seller is exactly the same in Three-Buyer as in Two-Buyer
	(\autoref{fig:two-buyer-proc-seller-mpst4}). In particular, Seller does not
	know that it communicates with Buyer3 instead of Buyer2. Thus, delegation is
	hidden from each role not involved. \qed
\end{example}

\subsubsection*{Generic global types}

By embedding global\slash local types as Scala types, Scala's built-in mechanism
of type parametrisation is readily available. This allows the programmer to
write \textit{generic global types} with type parameters for roles (common in
external-DSL-based MPST tools) and sub-protocols (novel of \tool).

\begin{figure}[t]\centering
	\begin{minipage}[t]{\linewidth*3/7-2mm}
		\begin{lstlisting}[language=scala, numbers=left]
type T[P <: Role, Q <: Role] =
  Com[P, Q, ((Propose,
    Loop["X",
      U[Q, P,
        U[P, Q, Recur["X"]]]]))]
		\end{lstlisting}
	\end{minipage}%
	\hfill%
	\begin{minipage}[t]{\linewidth*4/7-2mm}
		\begin{lstlisting}[language=scala, numbers=left, firstnumber=last]
type U[P <: Role, Q <: Role, G <: GType] =
  Com[P, Q, (
    (Accept,  Com[Q, P, ((Confirm, End))]),
    (Reject,  End),
    (Propose, G))]
		\end{lstlisting}
	\end{minipage}%
	
	\caption{Generic global types for Negotiation}
	\label{fig:auth-glob[]-mpst4}
\end{figure}

\begin{example}
	\label{exmp:nego[]}
	
	To alleviate the repetitive feel of \autoref{fig:nego-glob-mpst4},
	\autoref{fig:auth-glob[]-mpst4} shows generic global types that leverage type
	parameters. Type \scala!U! generically specifies the communication of an
	acceptance, rejection, or counter-proposal from \scala!P! to \scala!Q! (type
	parameters for roles), followed by \scala!G! (type parameter for a
	sub-protocol) in case of a counter-proposal; it can be instantiated twice to
	replace lines 5--7 and lines 8--10 in \autoref{fig:nego-glob-mpst4}. This is
	done in type \scala!T!, which generically specifies a role-parametric version
	of the whole \scalatype!S! in \autoref{fig:nego-glob-mpst4}. Thus, %
	\scala!T["A", "B"]! is equivalent to \scalatype!S! in
	\autoref{fig:nego-glob-mpst4}. \qed
\end{example}

\subsubsection*{Consistency}

\tool also supports explicit \textit{consistency checking} of sets of local
types. Details can be found in the \iftr{appendix
(\autoref{sect:cons})}{technical report \cite{techreport}}, as they are rather
technical\slash subtle. We do evaluate consistency checking times in
\autoref{sect:overv:perf}, though.


\section{Technical Details}
\label{sect:detail}

As \tool closely follows MPST theory, and as it uses unique parts of the
Scala type system, first, we summarise a few essential preliminaries
(\autoref{sect:detail:prelim}). Next, we describe our embedding of MPST into
Scala (\autoref{sect:detail:detail}).

This section focusses on the basic features of \autoref{sect:overv:basics}. It
allows us to keep the necessary background on MPST theory simple and succinct,
while still being able to explain the general ideas of the embedding into the
Scala type system in sufficient depth. The advanced features of
\autoref{sect:overv:merge} are based on more complex theoretical concepts, but
their embedding follows similar general ideas.

\subsection{Preliminaries}
\label{sect:detail:prelim}

\subsubsection*{MPST theory}

We summarise the theory behind classical MPST (\autoref{fig:mpst}):

\begin{description}
	\item[Global types, local types, and processes:] The syntax was defined and
	explained in \autoref{fig:relation} (right column) and \autoref{exmp:nego}.
	
\begin{figure}[t]
	\begin{minipage}{\linewidth}\centering
		$\begin{aligned}
			p \unbuf q \isa \famxx{t_i \pre G_i}{i \in I} \proj p &= pq \send \famxx{t_i \pre G_i \proj r}{i \in I}
		\\
			p \unbuf q \isa \famxx{t_i \pre G_i}{i \in I} \proj q &= pq \recv \famxx{t_i \pre G_i \proj r}{i \in I}
		\\
			p \unbuf q \isa \famxx{t_i \pre G_i}{i \in I} \proj r &= {\textstyle\bigsqcap} \famxx{G_i \proj r}{i \in I} \quad\text{if } r \notin \setx{p, q}
		\end{aligned}
		\qquad
		\begin{aligned}
			\one \proj r &= \one
		\\
			\recx{X} \pre G \proj r &= \recx{X} \pre (G \proj r)
		\\
			X \proj r &= X
		\end{aligned}$
		\caption{Projection in MPST theory}
		\label{fig:project}
	\end{minipage}
\end{figure}
	
	\item[Projection:] Let $G \proj r$ denote the projection of $G$ onto $r$; it is
	defined in \autoref{fig:project}.
	
	The projection of a communication yields a send if $r$ is the sender, a receive
	if $r$ is the receiver, or the full merge---denoted by $\sqcap$---of the
	projected branches otherwise (i.e., $r$ does not participate in the
	communication).
	
\begin{figure}[t]
	\begin{minipage}{\linewidth}\centering
	
		$\begin{gathered}
			\rulexxx{Send}{
				\varGamma \vdash e \typed t_j
			\text{ and }
				\varGamma, x \typed L_j \vdash P
			\text{, for some } j \in I
			}{
				\varGamma, x \typed pq \send \famxx{t_i \pre L_i}{i \in I} \vdash x \atx{pq} \send e \pre P
			}
		\qquad
			\rulexxx{Term}{
				\text{only $\checkmark$ local types in $\varGamma$}
			}{
				\varGamma \vdash \nil
			}
		\\[.5\baselineskip]
			\rulexxx{Recv}{
				\varGamma, x \typed L_i, x_i \typed t_i \vdash P_i \text{, for each } i \in I
			}{
				\varGamma, x \typed pq \recv \famxx{t_i \pre L_i}{i \in I}) \vdash x \atx{pq} \recv \famxx{x_i \isa t_i \pre P_i}{i \in I}
			}
		\qquad
			\rulexxx{Unfold}{
				\varGamma, x \typed L [\recx{X} \pre L / X] \vdash P
			}{
				\varGamma, x \typed \recx{X} \pre L \vdash P
			}
		\end{gathered}$
		\caption{Type checking in MPST theory (excerpt)}
		\label{fig:typing}
	\end{minipage}
\end{figure}
	
	\item[Type checking:] Let $\varGamma \proves[] P$ denote well-typedness of $P$
	in typing environment $\varGamma$; it is defined in \autoref{fig:typing}. Rule
	\autorefrule{Send} states that a send from $p$ to $q$ in $x$ is well-typed when
	the local type of $x$ specifies a send, $e$ is well-typed by $t_j$, and $P$ is
	well-typed after setting the local type of $x$ to $L_j$ in the typing
	environment, for some $j$. Rule \autorefrule{Recv} states that a receive from
	$p$ to $q$ in $x$ is well-typed when the local type of $x$ specifies a receive,
	and $P_i$ is well-typed after setting the local type of $x$ to $L_i$ in the
	typing environment, for each $i$. Thus, there is asymmetry: for sending, only
	one send specified must be implemented, but for receiving, each receive
	specified must be implemented.
	
	\item[Central theorem:] Static well-typedness implies dynamic safety and liveness.
\end{description}

\subsubsection*{Match types in Scala}

The main feature of the Scala type system that we take advantage of in \tool is
\textit{match types}. We explain it with an example. Suppose that we need
to write a function to convert \scala^Int^s and \scala^Boolean^s:
\medskip\begin{lstlisting}[language=scala]
type IntOrBoolean = Int | Boolean // type alias for a union type
def convert(x: IntOrBoolean): IntOrBoolean = x match {
  case i: Int => i == 1; case b: Boolean => if b then 1 else 0 }
\end{lstlisting}\medskip
\noindent However, return type
\scala^IntOrBoolean^ is not precise enough. For instance, the Scala
compiler fails to prove that \scala^convert(5) && false^ is safe, as
it cannot infer that \scala^convert(5)^ is \scala^Boolean^. 
What is missing, is a relation between the actual type of \scala^x^
(e.g., \scala^Int^) and the return type (e.g., \scala^Boolean^). Match types
 define such relations.

\begin{enumerate}
	\item First, we redefine the signature of \scala^convert^ as follows:
\medskip\begin{lstlisting}[language=scala]
def convert[T <: IntOrBoolean](x: T): Convert[T] = ... // same as before
\end{lstlisting}\medskip
	Thus, we introduce a type parameter \scala^T^ (subtype of
	\scala^IntOrBoolean^) and declare \scala^x^ to be \scala^T^.
	Also, we declare the return value to be of match type
	\scala^Convert[T]^. 
	
	\item Next, the idea is to define \scala^Convert[T]^ in such a way that the
	relation between the actual type of \scala^x^ and the return type can be
	inferred, as follows:
\medskip
\begin{lstlisting}[language=scala]
type Convert[T] = T match { case Int => Boolean; case Boolean => Int }
\end{lstlisting}
\medskip
	\noindent The Scala compiler \textit{reduces} every occurrence of
	\scala^Convert[T]^ to \scala^Int^ or \scala^Boolean^, depending on the
	instantiation of \scala^T^ (e.g., \scala^Convert[Int]^ is reduced to
	\scala^Boolean^).
	
	\item Last, for instance, the Scala compiler correctly succeeds/fails to
	type-check \scala^convert(5) && false^ (safe) and %
	\scala^convert(5) && 6^ (unsafe).
\end{enumerate}
\noindent Thus, match types are a ``lightweight form of dependent
typing''~\cite{DBLP:journals/pacmpl/BlanvillainBKO22}, to perform ``type-level
programming''. In the remainder, we use the following built-ins:
\begin{itemize}
	\item \scala!Head[(T1, ..., Tn)]! and \scala!Last[(T1, ..., Tn)]! reduce to
	\scala!T1! and \scala!Tn!.
	
	\item \scala!Map[(T1, ..., Tn), F]! reduces to \scala!(F[T1], ..., F[Tn])!. We
	note that \scala!F! can be a \textit{type lambda} of the form %
	\scala![X] => ... /* do something with X */!.
\end{itemize}

\subsection{Embedding MPST into Scala}
\label{sect:detail:detail}

\subsubsection*{Global types, local types, processes}

As explained in \autoref{sect:overv:basics}, and as shown in
\autoref{fig:relation}, global types and local types are implemented as classes,
while processes are implemented as methods of class \scala!Local!. The
communication infrastructure for processes is based on concurrent queues.
However, a transport layer for distributed processes is also possible
(orthogonal concern).

\subsubsection*{Projection}

\begin{figure}[t]
	\begin{minipage}{\linewidth}
		\begin{lstlisting}[language=scala, numbers=left]
type Proj[G, R] = G match
  case End          => End
  case Com[R, q, b] => Send[R, q, Map[b, [E] =>> (Head[E], Proj[Last[E], R])]]
  case Com[p, R, b] => Recv[p, R, Map[b, [E] =>> (Head[E], Proj[Last[E], R])]]
  case Com[_, _, b] => MergAll[Map[b, [E] =>> Proj[Last[E], R]]
  case Loop[x, g]   => Loop[x, Proj[g, R]]
  case Recur[x]     => Recur[x]
		\end{lstlisting}
		\caption{Projection in \tool}
		\label{fig:mpst4-project}
	\end{minipage}
\end{figure}

\autoref{fig:mpst4-project} shows match type \scala!Proj!. It is used to have
the Scala compiler fully automatically and statically compute local types (e.g.,
line 2 in \autoref{fig:nego-proc-bob-mpst4}).

Match type \scala!Proj! has two type parameters: a global type \scala!G! and a
role \scala!R! (cf. $G \proj r$). To reduce \scala!Proj[G, R]!, the Scala
compiler matches \scala!G! to a global type constructor, and it produces a local
type \textbf{exactly as defined in \autoref{fig:project}}. By convention, lower
case letters in patterns are \textit{type variables}; they are bound to types as
part of the matching algorithm. For instance, on lines 3--5 in
\autoref{fig:mpst4-project}, \scala!b! is bound to a product type of the form %
\scala!((T1, G1), ..., (Tn, Gn))!, where each \scala!Ti! is a data type, and
each \scala!Gi! is a global type. When \scala!b! is passed to \scala!Map! on
lines 3--4, it is converted into %
\scala!((T1, Proj[G1, R]), ..., (Tn, Proj[Gn, R]))!. Alternatively, when
\scala!b! is passed to \scala!Map! on line 5, it is converted into %
\scala!(Proj[G1, R], ..., Proj[Gn, R])!, which is subsequently passed to
\scala!MergAll!; this is a helper match type that reduces to the full merge of
all local types in the product type.

\subsubsection*{Type checking}

\begin{figure}[t]
	\begin{minipage}{\linewidth}
		\begin{lstlisting}[language=scala, numbers=left]
class Local[R, L] private (val r: R, val net: Network) extends UseOnce:
  def send[Q, D](q: Q, d: D, f: SendCallback[Q, D, L]): Local[R, End] = ...
  def recv[P](p: P, fs: RecvCallbacks[P, L]): Local[R, End] = ...
  
  type SendCallback[Q, D, L] = L match
    case Send[R, q, b] => q match
      case Q => (Local[R, App[b, D]] => Local[R, End])
    case Loop[x, l] => SendCallback[Q, D, Substitute[l, L, x]]
  
  type RecvCallbacks[P, L] = L match
    case Recv[p, R, b] => p match
      case P => Map[b, [E] =>> (Head[E], Local[R, Last[E]]) => Local[R, End]]
    case Loop[x, l] => RecvCallbacks[P, Substitute[l, L, x]]
  
  ... // function loop and type LoopCallback
		\end{lstlisting}
		\caption{Type checking in \tool (excerpt)}
		\label{fig:mpst4-typing}
	\end{minipage}
\end{figure}

\autoref{fig:mpst4-typing} shows an excerpt of class \scala!Local! related to
type checking. The idea is to have the Scala compiler reduce match types
\scala!SendCallback! and \scala!RecvCallbacks! to fully automatically and
statically compute the \textit{expected types} of the callback arguments of
methods \scala!send! and \scala!recv!, given a local type $L$. The reduction
succeeds, and the actual callback argument is well-typed by the expected type,
\textit{if, and only if,} the communication action is well-typed by $L$
\textbf{exactly as defined in \autoref{fig:typing}}. Otherwise, the Scala
compiler reports an error. In this way, \tool implements the same MPST typing
rules as in \autoref{fig:typing} in terms of Scala match type reduction, and it
provides the same assurances (modulo the ``two provisos'';
\autopageref{provisos}):

\begin{itemize}
	\item if, at compile-time, each process is well-typed by its projection,
	
	\item then, at run-time, the session of all processes is safe and live,
	
	\item modulo linear usage of \scala!Local! objects (checked dynamically),
	
	\item modulo non-terminating\slash ex\-cep\-tion\-al behaviour (unchecked).
\end{itemize}

We now explain \scala!send! and \scala!recv!. Regarding \scala!send!,
\autoref{fig:typing} states that a send is well-typed if the local type specifies it directly (rule \autorefrule{Send}) or indirectly (rule
\autorefrule{Unfold}). These cases correspond precisely to the two cases in
\scala!SendCallback!:

\begin{itemize}
	\item Lines 6--7 state that a send is well-typed if the sender, receiver, and
	data type match the send of the local type \scala!L!, and if the callback is a
	function that consumes a \scala!Local! object, parametrised by the selected
	branch of \scala!L!, namely %
\scala!App[b, D]!. We note that %
	\scala!App[((T1, L1), ..., (Tn, Ln)),!\allowbreak\scala!Ti]! reduces to
	\scala!Li!.

	\item Line 8 states that a send is also well-typed when it is well-typed by the
	unfolding of the local type. We note that \scala!Substitute[L1, L2, X]! reduces to a version of \scala!L1! in which
	each occurrence of \scala!Recur[X]! is replaced with \scala!L2!.
\end{itemize}

Regarding \scala!recv!, similarly, \autoref{fig:typing} states that a receive is
well-typed if the local type specifies the receive directly (rule
\autorefrule{Recv}) or indirectly (rule \autorefrule{Unfold}). Due to the
asymmetry between sends and receives, \scala!SendCallback! (singular) reduces
to a single function type, while \scala!RecvCallbacks! (plural) reduces to a
product of function types, computed using \scala!Map!. Besides that, they follow
the same ideas.

\subsection{Evaluation and Discussion}
\label{sect:overv:perf}

\subsubsection{Compile-time performance}

To validate the practical feasibility of using \tool, we systematically measured
the type checking times during \textit{non-incremental} compilation of all
examples in \autoref{sect:overv:basics} and \autoref{sect:overv:merge}, as well
as twelve additional examples from the MPST literature
\cite{DBLP:journals/mscs/CoppoDYP16,DBLP:journals/pacmpl/ScalasY19,DBLP:conf/ecoop/ScalasDHY17} and the Scribble repository \cite{demos}.\footnote{Run-time performance (e.g., latency/throughput) depends on the transport mechanism for message passing, which is orthogonal to the contributions of this paper.} This is a representative set of protocols, previously developed by other researchers (including the protocols in our examples in \autoref{sect:overv:basics} and \autoref{sect:overv:merge}), of various sizes, that exercise all aspects of classical MPST theory.\footnote{\label{classical}That is, the theory as originally defined by Honda et al. \cite{DBLP:conf/popl/HondaYC08}, but presented in the more recent style of, e.g., Scalas--Yoshida \cite{DBLP:journals/pacmpl/ScalasY19}, including the full merge operator.}

To measure only the protocol-related type checking times, the processes
contained almost no computation code; just communication actions in compliance
with the protocol. The measurements were obtained using an Intel i7-8569U
processor (4 physical/4 virtual cores at 2.8 GHz) and 16 GB of memory, running
macOS 14.0, OpenJDK 18.0.2, and Scala 3.3.1. We ran the measurements with
consistency checking disabled and enabled, to be able to study the difference.

\begin{table}[t]
\hspace{-\baselineskip}
	\caption{Type checking times in milliseconds, reported as $\mu \pm \sigma$, where $\mu$ is the average (of $31$ measurements), and where $\sigma$ is the standard deviation}
	\label{tab}

	\newcommand{\timexx}[2]{$#1 \pm #2$ ms}
	\newcommand{\pad}[1][0]{\phantom{#1}}
	\centering
	\begin{tabular}{@{}l@{\quad}r@{\quad}r@{\quad}r@{}}
		\toprule
		\bf{protocol} & \begin{tabular}{@{}c@{}}\bf{type checking}\\without consistency\end{tabular} & \begin{tabular}{@{}c@{}}\bf{type checking}\\with consistency\end{tabular} & \textbf{difference}$^\dagger$
	\\	\midrule
		{Negotiation} (\autoref{exmp:nego3}) & \timexx{1,399}{118} & \timexx{1,254}{\pad36} & \timexx{-145}{125}
	\\	{Negotiation} (\autoref{exmp:nego[]}) & \timexx{1,299}{\pad85} & \timexx{1,240}{\pad22} & \timexx{-60}{\pad88}
	\\	{Two-Buyer} (\autoref{exmp:two-buyer})& \timexx{1,399}{\pad31} & \timexx{1,658}{\pad50} & \timexx{258}{\pad59}
	\\	{Three-Buyer} (\autoref{exmp:three-buyer}) & \timexx{1,489}{\pad57} & \timexx{1,728}{\pad50} & \timexx{239}{\pad76}
	\\	\midrule
		{Three-Buyer} \cite{DBLP:journals/mscs/CoppoDYP16} & \timexx{1,341}{\pad57} & \timexx{1,622}{\pad71} & \timexx{280}{\pad78}
	\\	\midrule
		{OAuth2 Fragment} \cite{DBLP:journals/pacmpl/ScalasY19} & \timexx{\pad[0,]713}{\pad24} & \ooalign{\phantom{\timexx{1,254}{36}}\cr\hfil\textit{inconsistent}\hfil} & \ooalign{\phantom{\timexx{1,254}{36}}\cr\hfil\textit{inconsistent}\hfil}
	\\	Rec. Two-Buyers \cite{DBLP:journals/pacmpl/ScalasY19} & \timexx{\pad[0,]775}{\pad24} & \ooalign{\phantom{\timexx{1,254}{36}}\cr\hfil\textit{inconsistent}\hfil} & \ooalign{\phantom{\timexx{1,254}{36}}\cr\hfil\textit{inconsistent}\hfil}
	\\	Rec. Map/Reduce \cite{DBLP:journals/pacmpl/ScalasY19} & \timexx{1,016}{\pad45} & \ooalign{\phantom{\timexx{1,254}{36}}\cr\hfil\textit{inconsistent}\hfil} & \ooalign{\phantom{\timexx{1,254}{36}}\cr\hfil\textit{inconsistent}\hfil}
	\\	MP Workers \cite{DBLP:journals/pacmpl/ScalasY19} & \timexx{\pad[0,]891}{\pad27} & \ooalign{\phantom{\timexx{1,254}{36}}\cr\hfil\textit{inconsistent}\hfil} & \ooalign{\phantom{\timexx{1,254}{36}}\cr\hfil\textit{inconsistent}\hfil}
	\\	\midrule
		Game \cite{DBLP:conf/ecoop/ScalasDHY17} & \timexx{1,095}{\pad35} & \timexx{1,338}{\pad29} & \timexx{243}{\pad46}
	\\	\midrule
		{Adder} \cite{demos} & \timexx{\pad[0,]763}{\pad21} & \timexx{\pad[0,]781}{\pad19} & \timexx{18}{\pad28}
	\\	{Booking} \cite{demos} & \timexx{1,099}{\pad35} & \ooalign{\phantom{\timexx{1,254}{36}}\cr\hfil\textit{inconsistent}\hfil} & \ooalign{\phantom{\timexx{1,254}{36}}\cr\hfil\textit{inconsistent}\hfil}
	\\	{Fibonacci} \cite{demos} & \timexx{\pad[0,]759}{\pad22} & \timexx{\pad[0,]779}{\pad17} & \timexx{20}{\pad28}
	\\	{HTTP} \cite{demos} & \timexx{1,703}{\pad41} & \timexx{1,838}{\pad77} & \timexx{134}{\pad88}
	\\	{Loan Application} \cite{demos} & \timexx{\pad[0,]879}{\pad48} & \timexx{1132}{\pad29} & \timexx{253}{\pad56}
	\\	{SMTP} \cite{demos} & \timexx{1,726}{\pad70} & \timexx{2079}{128} & \timexx{353}{146}
	\\	\bottomrule
	\end{tabular}

	\smallbreak\footnotesize
	$^\dagger$ The difference between type checking times \textit{without} consistency $\mu_1 \pm \sigma_1$ and \textit{with} consistency $\mu_2 \pm \sigma_2$ are reported as $\mu \pm \sigma$, where $\mu = \mu_2 - \mu_1$ and $\sigma = \sqrt{\sigma_1^2 + \sigma_2^2}$
\end{table}

\autoref{tab} shows the results, averaged over 31 runs per protocol. We make two
main observations. First, without consistency checks, the type checking times
seem sufficiently low for the usage of \tool to be practically feasible: less
than two seconds for the biggest protocol in our benchmark set (SMTP). Moreover,
our measurements were obtained using non-incremen\-tal compilation and, as such,
constitute an upper bound on the expected type checking delays when using
incremental compilation. Anecdotally, in our development environment (Visual
Studio Code 1.87 with the Metals 1.30 extension for Scala programming), when
using incremental compilation, the type checking delays were significantly lower
(${<}100$ ms) than those in \autoref{tab}, and not disruptive at all.

Second, with consistency checks (i.e., five examples violate consistency; this
was expected), the results show that some overhead is added, but it does not
make the usage of \tool infeasible (${<}500$ ms). Also, when using incremental
compilation, the type checking delays continued to not get in the way.

\subsubsection{Experience}

The implementation of the benchmark set turned out to be, in its own right, a
validation activity to experience whether or not the type checker catches all
mistakes in practice. This is because, until a protocol implementation is
finished, it does not comply with the specification yet. Thus, all until the
end, the type checker reports errors to point out missing pieces. This guidance
by the type checker effectively prevented us from making unintended programming
mistakes, especially when writing the implementations of HTTP and SMTP (which
are the more complicated protocols in our benchmark set). It would be
interesting to try to reproduce these anecdotal findings in a larger user study.

\subsubsection{Expressiveness}

Our benchmark set shows that \tool is feature-com\-plete relative to classical
MPST theory,$^\text{\ref{classical}}$ with full merging (e.g., the OAuth2
fragment requires full merge), as intended. Moreover, while the ability to write
\textit{generic} global types does not add expressive power in the formal sense,
it enables better \textit{reuse} of global types and serves as an
abstraction/composition mechanism: it allows large protocols to be split into
separate smaller sub-protocols---specified as generically as possible to
maximise the opportunity for reuse---which can then be ``invoked'' from each
other with concrete arguments. Such generic sub-protocols can also be packaged
into libraries and shared between projects.

\section{Conclusion}
\label{sect:concl}

\subsubsection*{Related work}

Closest to our approach in this paper is the work by Imai et al.
\cite{DBLP:conf/ecoop/ImaiNYY19}. They developed an internal DSL in OCaml to
specify protocols and verify processes based on MPST. However, their tool does
\textit{not} support all key aspects of classical MPST as in \autoref{fig:mpst}:
it supports only binary choices instead of $n$-ary choices (e.g.,
\autoref{exmp:nego}, which has ternary choices, is not supported), and it is not
fully automatic (i.e., Imai et al. require the programmer to manually write
extra protocol-specific code to project global types). In contrast, \tool
supports $n$-ary choices and is fully automatic.

Another related tool is the Discourje library \cite{DBLP:conf/tacas/HamersJ20},
which offers an MPST-based internal DSL in Clojure. However, Discourje does all
verification dynamically, whereas \tool performs all verification statically
up-to linearity.

\begin{table}[t]
	\vspace{-\baselineskip}
	\caption{Comparison of MPST tools for Scala}
	\label{tab:scala-mpst}
	
	\centering
	\begin{tabular}{@{}l@{\quad}c@{\quad}c@{\quad}c@{\quad}c@{}}
		\toprule
		&\bf DSL &\bf projection &\bf interpretation &\bf encoding
	\\	\midrule
		Scribble-Scala \cite{DBLP:conf/ecoop/ScalasDHY17} & external & syntactic & FSMs & \lchannels
	\\	Pompset \cite{DBLP:conf/ecoop/CledouEJP22} & external & syntactic & pomsets & vanilla Scala
	\\	Teatrino \cite{DBLP:conf/ecoop/BarwellHY023} & external & syntactic & -- & Effpi
	\\	Oven \cite{DBLP:conf/issta/FerreiraJ23} & external & semantic & FSMs & vanilla Scala
	\\	\midrule
		\tool & internal & syntactic & -- & vanilla Scala
	\\	\bottomrule
	\end{tabular}
\end{table}

There are four existing tools to combine MPST with Scala:
\textit{Scribble-Scala} \cite{DBLP:conf/ecoop/ScalasDHY17}, \textit{Pompset}
\cite{DBLP:conf/ecoop/CledouEJP22}, \textit{Teatrino}
\cite{DBLP:conf/ecoop/BarwellHY023}, and \textit{Oven}
\cite{DBLP:conf/issta/FerreiraJ23}. \autoref{tab:scala-mpst} summarises the
differences:

\begin{itemize}
	\item \textit{DSLs to specify protocols as global types:} Scribble-Scala,
	Pompset, and\\ Teatrino are based on the external DSL of Scribble, while Oven
	is based on an external DSL for regular expressions.
	
	\item \textit{Projection of global types:} Scribble-Scala, Pompset, and
	Teatrino apply the classical \textit{structural} projection operator (defined
	in terms of the syntax of global types; \autoref{sect:detail:detail}), while
	Oven applies a non-classical \textit{behavioural} projection operator (defined
	in terms of the operational semantics of global types). The latter has
	additional expressive power to support the usage of regular expressions as
	global types \cite{DBLP:conf/ecoop/JongmansF23}.
	
	\item \textit{Interpretation of local types:} Different from
	\autoref{fig:apigen}, Pompset uses \textit{partially-ordered multisets} instead
	of FSMs as an intermediate operational model, while Teatrino directly encodes
	local types as APIs in Scala.
	
	\item \textit{Encoding as APIs:} The APIs generated by Scribble-Scala and
	Teatrino are built on top of the existing libraries \lchannels and Effpi
	(discussed in more detail below), while Pompset and Oven do not rely on such
	existing libraries.
\end{itemize}

Besides these existing tools to combine \underline{\smash{multiparty}} session
typing with Scala (including global types and projection), there also exist
libraries to combine \underline{\smash{binary}} session typing with Scala
(excluding global types and projection), namely \lchannels
\cite{DBLP:conf/ecoop/ScalasDHY17} and \textit{Effpi}
\cite{DBLP:conf/pldi/ScalasYB19}. Conceptually, as \tool targets multiparty
instead of binary, it is not really comparable to \lchannels and Effpi.
Technically, moreover, \lchannels and Effpi do not use match types.

\subsubsection*{Future work}

Many extensions of MPST theory have been proposed. We are keen to explore which
of them can be incorporated in \tool using match types. For instance, an
important feature that we believe is compatible with \tool and match types is
\textit{parameterised MPST with indexed roles} as developed by Castro et al.
\cite{DBLP:journals/pacmpl/CastroHJNY19}. Another feature that seems
representable using match types, is \textit{MPST with refinements} along the
lines of Zhou et al. \cite{DBLP:journals/pacmpl/00020HNY20}. In contrast, a
feature that seems prohibitively difficult to incorporate, is \textit{timed
MPST} \cite{DBLP:conf/concur/BocchiYY14}: match types seem unsuitable to
statically offer real-time guarantees.

\section*{Data Availability Statement}

The artifact is available on Zenodo \cite{artifact}. It contains: (1) \tool; (2) the examples in the paper; (3) reproduction instructions for our
evaluation.

\bibliographystyle{splncs04}
\bibliography{main}

\clearpage
\appendix

\ifdefined\tr

\section{Screenshot}
\label{sect:screen}

\begin{figure}[t]
	\includegraphics[width=\linewidth]{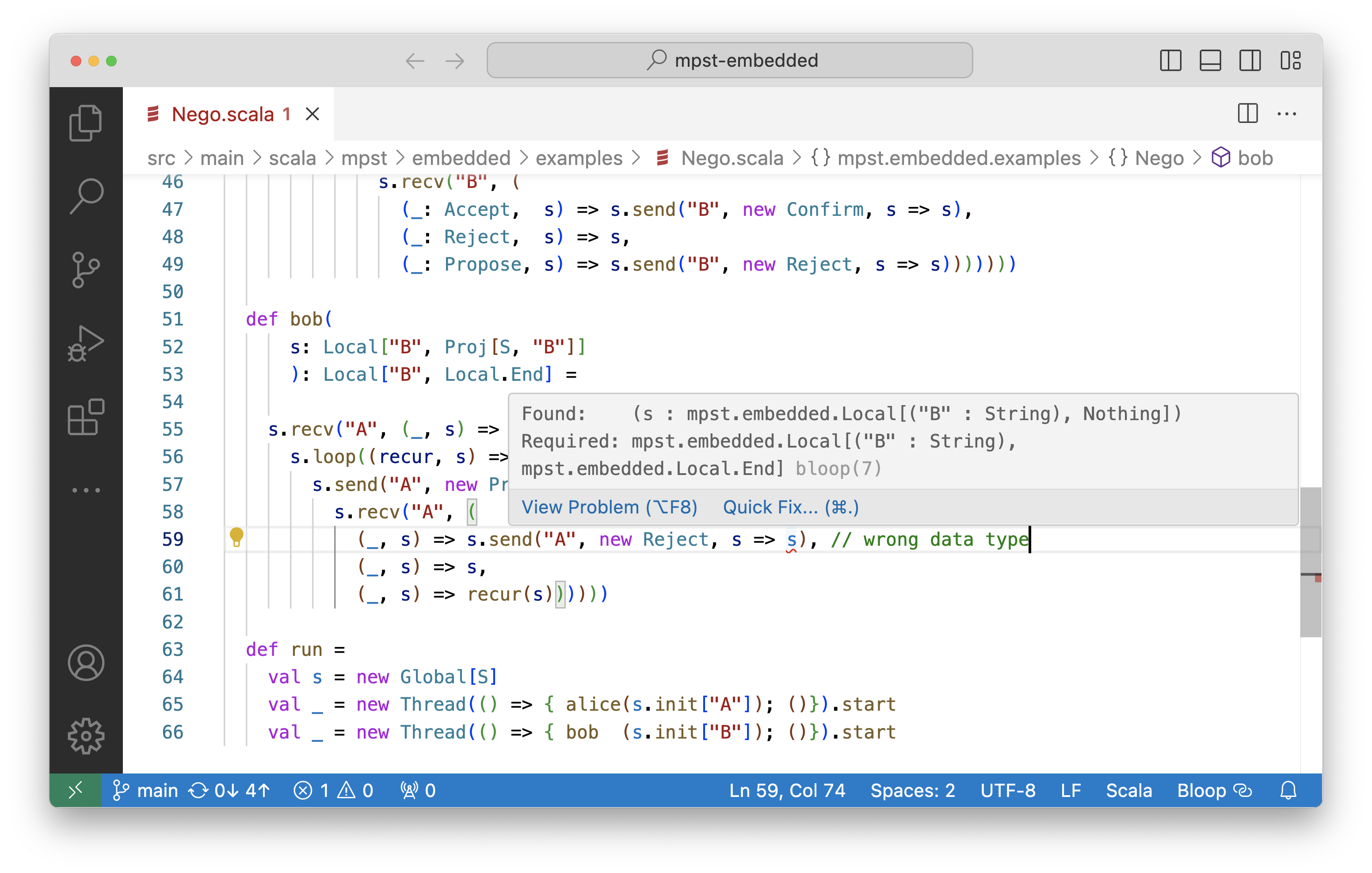}
	
	\caption{Screenshot of an error message in Visual Studio Code}
	\label{fig:screen}
\end{figure}

\autoref{fig:screen} shows a screenshot to demonstrate how errors at
compile-time manifest when using Visual Studio Code with the Metals extension
for Scala programming; it exemplifies the ``wrong data type'' protocol violation
in \autoref{exmp:errors}.

In general, when a call to \scala!send!/\scala!recv! violates the protocol, an
error is reported \textit{at the callback argument} of the call (e.g., indicated
by the red squiggly line in \autoref{fig:screen}, line 59). As it is currently
not possible in Scala to generate custom type error messages at
compile-time,\footnote{\url{https://github.com/lampepfl/dotty/pull/7951}} our
error messages for protocol violations are formulated in terms of Scala types
(as in \autoref{fig:screen}). As a result, \tool gives static feedback about
\textit{that} the protocol is violated, and \textit{where} in the code, but not
exactly \textit{how}; to determine the latter, some experience with \tool is
needed to be able to map Scala type error messages to protocol violations.

\section{Consistency Checking}
\label{sect:cons}

For well-typedness at compile-time to imply safety and liveness at run-time, in
classical MPST, each global type is assumed to be
\textit{consistent}.\footnote{Informally, consistency means that, for each pair
of roles $r_1$ and $r_2$, projections $G \proj r_1$ and $G \proj r_2$ are
inductively \textit{dual} relative to $r_2$ and $r_1$: if $G \proj r_1$
specifies a send to $r_2$, then $G \proj r_2$ specifies a receive from $r_1$,
and vice versa, recursively. As a result, neither sends without receives, nor
receives without sends, can happen (i.e., all sends and receives are properly
paired), so communication deadlocks cannot occur.} However, Scalas and Yoshida
showed that global type consistency is actually not guaranteed
\cite{DBLP:journals/pacmpl/ScalasY19}. For this reason, \tool supports explicit
consistency checks via type \scala!Consistent!: the Scala compiler reduces
\scalam!Consistent[$G$]! to boolean literal type \scalatype!true! if, and only
if, $G$ is consistent.

\begin{figure}[t]\centering
	\begin{minipage}[t]{\linewidth}
		\begin{lstlisting}[language=scala, numbers=left]
type S =
  Com["S", "C", (
    (Login,
      Com["C", "A", ((Password,
        Com["A", "S", ((Auth,
          End))]))]),
    (Cancel,
      Com["C", "A", ((Quit,
        End))]))]

val _ = true: Consistent[S]
		\end{lstlisting}
		\caption{Global type for Authorisation}
		\label{fig:auth-glob-mpst4}
	\end{minipage}%
\end{figure}

\begin{example}
	The Scala compiler reduces \scalam!Consistent[S$/$T]! to
	\scalatype!true! for each \scala!S!/\scala!T! in
	\autoref{fig:nego-glob-mpst4} (Negotiation), \autoref{fig:two-buyer-glob-mpst4}
	(Two-Buyer), and \autoref{fig:three-buyer-glob-mpst4} (Three-Buyer). \qed
\end{example}

\begin{example}
	\label{exmp:auth} The \textit{Authorisation} protocol, originally defined in
	the MPST literature by Scalas--Yoshida \cite{DBLP:journals/pacmpl/ScalasY19}
	(based on OAuth 2.0 \cite{rfc6749}), consists of roles \textit{Service}
	\textit{Client}, and \textit{AuthorisationServer}. First, a login request or
	cancellation is communicated from Service to Client. Next:
	
	\begin{itemize}
		\item In case of a login request, a password is communicated from
		Client to AuthorisationServer. Next, an authorisation status flag is communicated from
		AuthorisationServer to Service.
		
		\item In case of a cancellation, a quit signal is communicated.
	\end{itemize}
	
	\noindent \autoref{fig:auth-glob-mpst4} shows a global type for Authorisation,
	plus consistency check. The Scala compiler reduces \scala!Consistent[S]! to
	\scalatype!false!, reported as an error on line 9. \qed
\end{example}

Consistency checking and projection are decoupled in \tool: to offer
flexibility, we allow processes to be type-checked against the projections of
inconsistent global types. This is because consistency is sound (sufficient) but
not complete (necessary): there are plenty of inconsistent global types that are
nevertheless ``good enough'' for well-typedness to imply safety and liveness
\cite{DBLP:journals/pacmpl/ScalasY19}. However, if a programmer uses an
inconsistent global type to verify processes, it becomes their responsibility to
check that the global type is indeed ``good enough'' (e.g., using type-level
model checking \cite{DBLP:journals/pacmpl/ScalasY19}, outside of \tool).

\begin{example}
	Type \scalatype!S! in \autoref{fig:auth-glob-mpst4} is inconsistent but ``good
	enough'' \cite{DBLP:journals/pacmpl/ScalasY19}. \qed
\end{example}

\fi

\end{document}

%% file: commands.tex
\newcommand{\datax}[2][\dataxdefault]{\smash{\texttt{{\renewcommand{\ }{\phantom{x}}\upshape#1\fontdimen2\font=.375em#2}}}}

\newcommand{\proves}[1][\dict]{\mathrel{\provessym[#1]}}
\newcommand{\provessym}[1][\dict]{{\vdash_{#1}}}

\newcommand{\typed}{\mathrel{\typedsym}}
\newcommand{\typedsym}{{:}}

\newcommand{\rolex}[2][\rolexdefault]{{\pmb{\text{\upshape#1\texttt{#2}}}}}

\newcommand{\send}{\text{\upshape\ooalign{\phantom{?}\cr\hfil!\hfil\cr}}}
\newcommand{\recv}{\text{\upshape?}}

\newcommand{\proj}[1][]{\mathbin{\projsym[#1]}}
\newcommand{\projsym}[1][]{{\upharpoonright_{#1}}}

\newcommand{\typeannotx}[1]{{\color{teal}\typedsym#1}}

\newcommand{\recvx}[2][]{\recv(#2\IfStrEq{#1}{}{}{\typeannotx{#1}})}

\newcommand{\dict}{\smash{\mathfrak{D}}}

\usepackage{nicefrac}
\usepackage{centernot}

\newcommand{\nterm}[1][\dict]{{{}\mathrel{\centernot{\downarrow}}}}
\newcommand{\ntermsym}[1][\dict]{{\centernot{\downarrow}}}

\newcommand{\inxx}[3][]{#2(#3\ifthenelse{\equal{#1}{}}{}{{{\color{black}\isa#1}}})}
\newcommand{\nil}{\mathbf{0}}
\newcommand{\pre}{\mspace{1mu}{.}\mspace{1mu}}

\newcommandtwoopt{\loopx}[3][][]{\loopsym[#1]\ifthenelse{\equal{#2}{}}{}{(#2)}\mskip\medmuskip#3}
\newcommand{\loopsym}[1][]{\smash{\mathbf{loop}}_{#1}}
\newcommandtwoopt{\recur}[2][][]{\mathbf{recur}_{#1}\ifthenelse{\equal{#2}{}}{}{(#2)}}

\newcommand{\unbuf}{\mspace{1mu}{\unbufsym}\mspace{1mu}}
\newcommand{\unbufsym}{{\rightarrowtriangle}}
\newcommand{\isa}{\mspace{1mu}\typedsym\mspace{1mu}}
\newcommand{\one}{\smash{\checkmark}}

\newcommand{\recx}[1]{\upmu#1}

\newcommand{\prewide}{\mskip\medmuskip{.}\mskip\medmuskip}

\newcommandtwoopt{\rollsym}[2][][]{{\prec_{#1}^{#2}}}

\newcommand{\tool}{{\small\texttt{mpst.\allowbreak embedded}}\xspace}
\newcommand{\lchannels}{{\small\texttt{lchannels}}\xspace}

\newenvironment{tree}{%
	\begin{tikzpicture}[remember picture, anchor=base west, baseline={(0,0)}, inner sep=0pt, line width=0pt]%
		\newcommand{\holex}[1]{{\tikz{\coordinate (##1)}}}%
		\newcommand{\branchx}[2][0,0]{\node at (##1) {\ensuremath{##2}};}%
		\newcommandtwoopt{\branchxx}[4][0,0][]{\node[##2] at (##1) {\ensuremath{\branchherexx{##3}{##4}}};}%
		\newcommandtwoopt{\branchxxx}[5][0,0][]{\node[##2] at (##1) {\ensuremath{\branchherexxx{##3}{##4}{##5}}};}%
		\newcommand{\branchherexx}[2]{{\left\{\begin{aligned}{}&##1\\&##2\end{aligned}\!\right.}}%
		\newcommand{\branchherexxx}[3]{{\left\{\begin{aligned}{}&##1\\&##2\\&##3\end{aligned}\!\right.}}%
}{%
	\end{tikzpicture}%
}

\newcommand{\atx}[1]{\llbracket#1\rrbracket}

\lstdefinelanguage{scrib}{%
  keywords=[1]{choice,at,or,local,from,to,rec,continue,par,and,end,data,global,protocol,do,role},
  keywordstyle=[1]\color{purple},
  keywords=[2]{Accept,Confirm,Reject,Propose,Int},
  keywordstyle=[2]\color{teal},
  morecomment=[l]{//},
  morecomment=[s]{/*}{*/},
  commentstyle=\color{gray},
  mathescape=true,
}

\lstdefinelanguage{scala}{%
  alsoletter={@},
  keywords=[1]{new, if, then, else, class, return, case, match, import, type, val, trait, while, do, var, def, extends, throw, true, false, throw, private},
  keywordstyle=[1]\color{purple},
  keywords=[2]{Cup, Cap, Consistent, Loop, End, Recur, Map, Fold, Head, Last, Proj, Com, Merg, Tuple, S, T, T1, Tn, Ti, Ln, Li, F, Z, X, R, U, P, Q, Login, Password, Auth, Cancel, Role, Type, G, G1, Gn, Gi, D, E, L, L1, L2, S@B, S@S, Delegate, T@S, Merge, Send, Recv, Accept, Reject, Propose, Confirm, SendCallback, RecvCallbacks, Substitute, App, Network, UseOnce, State1, State2, State3, State4, State5, State6, S1, S2, S3, S4, S5, S6, Ok, Quit, Thread, String, Int, Boolean, IntOrBoolean, Convert, Date, LocalSession, GlobalSession, Project, Local, Proj, GType, Global, MergAll},
  keywordstyle=[2]\color{teal},
  morecomment=[l]{//},
  morecomment=[s]{/*}{*/},
  commentstyle=\color{gray},
  morestring=[b]",
  stringstyle=\color{black},
  mathescape=true,
}

\newcommand{\scribm}{\lstinline[language=scrib,mathescape=true,basicstyle=\ttfamily\small]}

\newcommand{\scala}{\lstinline[language=scala,mathescape=false,basicstyle=\ttfamily\small]}
\newcommand{\scalam}{\lstinline[language=scala,mathescape=true,basicstyle=\ttfamily\small]}
\newcommand{\scalatype}{\lstinline[language=scala,basicstyle=\ttfamily\small,keywordstyle={[1]\color{teal}},stringstyle={\color{teal}}]}

\usepackage{environ}
\NewEnviron{listgather*}{%
	\begin{gather*}
		\hspace{24.5pt}\hbox to \linewidth {\hfil$\begin{gathered}[t]
			\BODY
		\end{gathered}$\hfil}
	\end{gather*}%
}

\usetikzlibrary{decorations.pathreplacing,calligraphy}

\usetikzlibrary{calc}

\tikzstyle{state} = [inner sep=.5mm, outer sep=.25mm, circle, fill, line width=0pt]
\tikzstyle{final} = [inner sep=.75mm, outer sep=.25mm, circle, fill=none, draw]
\tikzstyle{trans} = [-stealth, rounded corners=1mm]
\tikzstyle{bitrans} = [stealth-stealth, rounded corners=1mm]
\tikzstyle{label} = [inner sep=0pt, line width=0pt, font=\scriptsize]
\tikzstyle{label-left} = [label, anchor=base east, xshift=-1mm, yshift=-.25ex]
\tikzstyle{label-right} = [label, anchor=base west, xshift=1mm, yshift=-.25ex]

\usetikzlibrary{arrows.meta}
\usetikzlibrary{decorations.pathmorphing}

\tikzstyle{object} = [font=\footnotesize]
\tikzstyle{wred} = [double, -{Implies}]
\tikzstyle{wbisim} = [double, decorate, decoration={snake, amplitude=.25mm, segment length=2mm}]

\newcommand{\autorefrule}[1]{\ensuremath{\text{\hyperlink{\detokenize{#1}}{[\textsc{#1}]}}}}

\newcommand{\famxx}[2]{\setx{#1}_{\smash{#2}}}

\newcommand{\setx}[1]{\{#1\}}

\let\primenosmash\prime
\let\astnosmash\ast
\let\dagnosmash\dag
\let\ddagnosmash\ddag
\let\Snosmash\S
\let\Pnosmash\P
\renewcommand{\prime}{{\smash{\primenosmash}}}
\renewcommand{\ast}{{\smash{\astnosmash}}}
\renewcommand{\dag}{{\smash{\dagnosmash}}}
\renewcommand{\ddag}{{\smash{\ddagnosmash}}}
\renewcommand{\S}{{\smash{\Snosmash}}}
\renewcommand{\P}{{\smash{\Pnosmash}}}

\newcommand{\dfracfrac}[2]{\text{%
  \rlap{\ensuremath{\dfrac{\phantom{#1}}{#2}}}%
  \raisebox{2pt}{\ensuremath{\dfrac{#1}{\phantom{#2}}}}%
}}

\newcommand{\rulexxx}[3]{\dfrac{#2}{#3}\ifthenelse{\equal{#1}{}}{}{\, \text{\upshape\scriptsize[\hypertarget{\detokenize{#1}}{\textsc{#1}}]}}}
\newcommand{\corulexxx}[3]{\dfracfrac{\begin{gathered}#2\end{gathered}}{\begin{gathered}#3\end{gathered}}\ifthenelse{\equal{#1}{}}{}{\, \text{\upshape\scriptsize[\hypertarget{\detokenize{#1}}{\textsc{#1}}]}}}

\newcommandtwoopt{\BLOCK}[3][{[}][{]}]{%
	\begingroup%
				\left#1\begin{gathered}#3\end{gathered}\right#2%
	\endgroup%
}

\newcommandtwoopt{\BLOCKt}[3][{[}][{]}]{%
	\begingroup%
				\left#1\begin{gathered}[t]#3\end{gathered}\right#2%
	\endgroup%
}

\NewEnviron{hide}{}
\let\oldproof\proof
\let\oldendproof\endproof

\newcommand{\hideproofs}{%
	\let\proof\hide%
	\let\endproof\endhide}
	
\newcommand{\showproofs}{%
	\let\proof\oldproof%
	\let\endproof\oldendproof}

\ifdefined\forceshowproofs
  \let\hideproofs\showproofs
\fi
\ifdefined\forcehideproofs
  \let\showproofs\hideproofs
\fi
